\documentclass[floatfix,prx,aps,showpacs, twocolumn]{revtex4-1}

\usepackage{graphicx,amsmath,amssymb,color,ulem}
\usepackage{nicefrac}
\usepackage{multirow, makecell, ragged2e}
\usepackage[titletoc,title]{appendix}
\usepackage{bm}
\usepackage{float,lipsum}
\usepackage{algpseudocode}
\usepackage{algorithm}

\newcommand{\be}{\begin{equation}}
\newcommand{\ee}{\end{equation}}

\newcommand{\cH}{\mathcal{H}}

\renewcommand{\vec}[1]{\bm{#1}}

\newcommand{\dd}{\mathrm{d}}
\newcommand{\unit}[1]{\vec{\hat{ #1}}}

\begin{document}

\title{Phase diagram of superconductivity in the integer quantum Hall regime}
\author{Jonathan Schirmer$^1$, C. -X. Liu$^{1, 2}$, J. K. Jain$^1$}
\affiliation{$^1$Department of Physics, 104 Davey Lab, The Pennsylvania State University, University Park, Pennsylvania 16802 \\
$^2$Department of Physics, Princeton University, NJ 08544, USA}
\begin{abstract}
An interplay between pairing and topological orders has been predicted to give rise to superconducting states supporting exotic emergent particles, such as Majorana particles obeying non-Abelian braid statistics. We consider a system of spinless electrons on a Hofstadter lattice with nearest neighbor attractive interaction, and solve the mean-field Bogoliubov-de Gennes equations in a self-consistent fashion,
 leading to gauge invariant solutions that display a rich phase diagram as a function of the chemical potential, magnetic field and the interaction. As the strength of the attractive interaction is increased, the system first makes a transition from a quantum Hall phase to a skyrmion lattice phase that is fully gapped in the bulk but has topological chiral edge current, characterizing a topologically non-trivial state. This is followed by a vortex phase in which the vortices carrying Majorana modes form a lattice; the spectrum contains a low-energy Majorana band arising from the coupling between neighboring vortex-core Majorana modes but does not have chiral edge currents. For some parameters, a dimer vortex lattice occurs with no low energy Majorana band. The experimental feasibility and the observable consequences of skyrmions as well as Majorana fermions are indicated.
\end{abstract}

\graphicspath{{./figures/}}

\maketitle

It has been nearly a century since the discovery of quantum mechanics, and yet some of its fundamental consequences are still being uncovered to this day. In particular, the role of topology in the quantum description, even for single-particle physics, had been largely ignored until the discovery of the integer quantum Hall effect (IQHE) \cite{klitzing1980new}, which has led to the development of topological band theory~\cite{Thouless1982}. In addition, a conceptually distinct but practically related development was that of quantum statistics in two-dimensional (2D) systems, 
which rely on the notion that two successive exchanges of a pair of particles in 2D space need not be an identity operation, as must be the case in three or higher dimensions. This fact leads to the possibility of new types of quasiparticles, called anyons, which generalize the quantum statistics of bosons and fermions \cite{Leinaas77,wilczek1982magnetic,Wu1984}.

Superconductivity (SC), when combined with non-trivial band topology such as
that found in the IQHE, can yield Majorana zero modes (MZMs), which are non-Abelian Ising anyons \cite{majorana1937,read2000,Ivanov01,Kitaev01,stone2006,Fu08,Wilczek2009,qi2010chiral,Sau2010,leijnse2012,beenakker2013,alicea2015,Elliott15}. Moreover, when interacting topological states, such as the fractional quantum Hall effect (FQHE) \cite{Tsui82}, are combined with SC, even richer and more exotic classes of quasiparticles, for example parafermions and Fibonacci anyons,
are thought to result. Besides being of
fundamental scientific interest, anyons (particularly non-Abelian anyons) have played a
role in proposals to engineer fault-tolerant topological quantum computation \cite{kitaev03,freedman03,sarma2006,stone2006,tewari2007,nayak08,Sau2010,alicea2012,pachos2012,stanescu2013,beenakker2013,mong2014,Jain20b}.

Previous works investigating SC in quantum Hall (QH) systems have focused on proximity-coupled superconductivity, so that the SC pairing potential has been treated as an external field, and has not been determined self-consistently \cite{Zocher2016,Lee17,jeon2019,Mishmash2019,chaudhary2020,chaudhary2021,vafek2001,PhysRevB.92.134519,pathak2021}. However, because the pairing potential is not gauge invariant (it transforms with charge $2e$), the choice of the form of the (proximity-induced) pairing potential 
must be specified together with the gauge. A natural choice for the pairing potential
in one gauge may not be natural in another. This may lead to ambiguities in physical observables such as the Chern number -- a manifestly gauge invariant quantity, since it counts edge modes of the system.

Previous works have studied the problem in the presence of a pairing field describing an Abrikosov vortex lattice -- the expected ground state for type-II superconductors subject to a magnetic field~\cite{franz2000,vafek2001,rosenstein2010ginzburg}.
These treatments do not allow for more exotic pairing order -- such as half quantum vortices (HQVs), skyrmions, and giant vortices -- that is possible in $p$-wave superconductor materials due to the multicomponent nature of the order parameter \cite{takigawa2001,babaev2002,li2009skyrmion,babaev2012,garaud2012,garaud2015,Babaev2017,Becerra2015,zhang2016}. Without a more complete, self-consistent treatment of the pairing order parameter, one cannot establish a priori what sort of order the ground state would adopt.

\begin{figure}
\includegraphics[width=.47\textwidth]{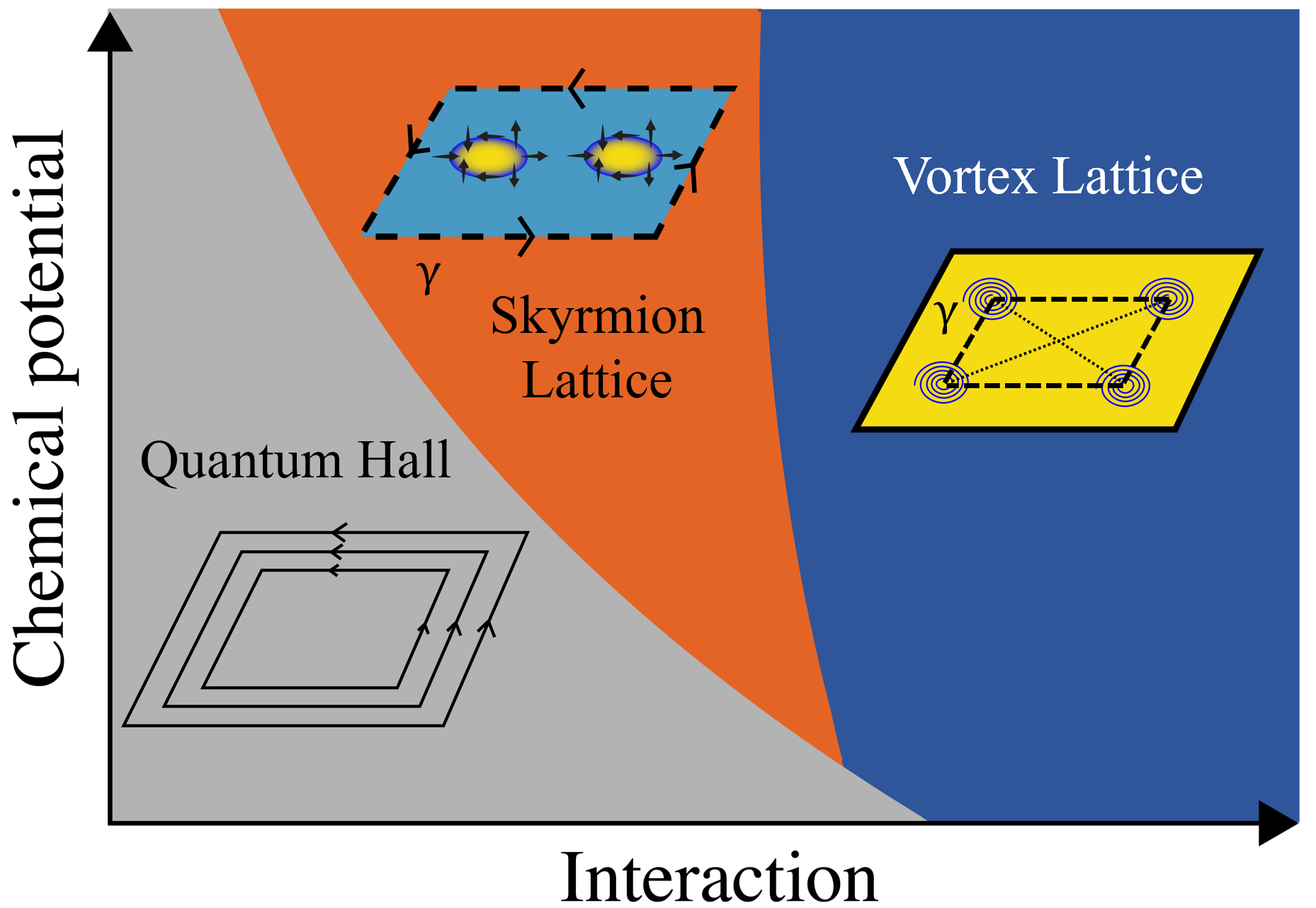}
\caption{A schematic phase diagram showing the phases of the model. The types of edge modes that the phases can support are indicated: the quantum Hall phase hosts chiral fermionic edge modes (solid lines); the skyrmion lattice phase hosts chiral Majorana edge modes (dashed lines labeled with $\gamma$); the vortex lattice phase either hosts fermionic edge modes, or no edge modes at all. In the vortex lattice phase, the Majorana excitations, which in a chiral $p$-wave supeprconductor are bound to vortices, hybridize between proximate vortices (depicted via dashed lines between vortices) and open a low energy spectral gap. In the skyrmion lattice phase the yellow and blue colors indicate different chiralities of the superconducting order parameter, and the arrows depict the direction of a pseudospin vector (defined in the text).} \label{schematic}
\end{figure}

Our approach in this paper is to solve the Bogoliubov-de Gennes (BdG) mean field equations self-consistently for a model of spinless fermions with attractive interactions on a square lattice, where self-consistency guarantees gauge invariance. We do not assume a form of the pairing potential, other than its being the result of a nearest neighbor attractive interaction, and therefore we treat the possible ground states on an equal footing. We use many random initial guesses for the self-consistent iterations in order to explore all topological sectors of solutions.  We are interested in the topological nature of the ground states, which is characterized by the (superconducting) Chern number of the BdG Hamiltonian (defined below). In particular, odd SC Chern number is an indicator for the non-Abelian, or topologically non-trivial, phase of 
superconductivity, which is of primary interest.

A schematic of the various phases following from our calculations is shown in Fig. \ref{schematic}; a more detailed phase diagram is given below. We find that upon increasing the strength of the attractive interaction $U$, the system transitions into a superconducting phase, as expected. However, typically, the 
ground state does not form an Abrikosov lattice of vortices, but instead
a lattice of skyrmions -- where a skyrmion texture is formed by the two component SC pairing functions and represents a domain of inverted chirality in a
chiral $p$-wave superconductor, 
with half quantum vortices bound along the domain wall.
In the skyrmion lattice phase, the SC Chern number is odd, and so the system hosts chiral Majorana edge modes. However, the bulk fermionic spectrum is gapped, implying that there are no MZMs in the bulk. Upon further increasing $U$, the system forms a square lattice of Abrikosov vortices. The MZMs located at the vortices hybridize into low-energy Majorana bands, but the ground state has an even SC Chern number, implying that the system is in a topologically trivial phase with no chiral Majorana edge modes.  In addition, at some isolated points in the parameter space, a lattice of vortex dimers may compete with the vortex lattice and become the ground state.
This phase, like the skyrmion phase, has chiral Majorana edge modes, because the SC Chern number is odd, but the bulk spectrum is gapped, indicating an absence of bulk MZMs.

\section*{Model}
We start with the interacting Hamiltonian, which is defined on a square lattice with unit lattice constant as
\begin{equation} \label{ham}
\begin{split}
&\cH = \cH_0 + \cH_I \\
&\cH_0=-\sum_{j,\delta}\Big(e^{iA_{j+\delta,j}}\hat{c}^\dagger_{j+\delta} \hat{c}_j+e^{iA_{j,j+\delta}}\hat{c}^{\dagger}_j \hat{c}_{j+\delta}\Big) -\mu \sum_j \hat{c}^\dagger_j \hat{c}_j \\
&\cH_I= -U\sum_{j,\delta}\hat{c}^\dagger_{j+\delta}\hat{c}^\dagger_j\hat{c}_j\hat{c}_{j+\delta}
\end{split}
\end{equation}
where $\hat{c}^{\dagger}_j$ ($\hat{c}_j$) creates (annihilates) a spinless fermion on site $j$, located at position $\vec{r}_j$, and $\hat{c}^{\dagger}_{j+\delta}$ ($\hat{c}_{j+\delta}$) creates (annihilates) a spinless fermion on a site which is a nearest-neighbor to site $j$, located at position $\vec{r}_j+\hat{\delta}$ where $\delta=x,y$.  The chemical potential is $\mu$, the interaction strength is $-U$ with $U>0$, and the external magnetic field is incorporated through complex hopping matrix elements with
phases $A_{j+\delta,j}=-A_{j,j+\delta}$.
We set the hopping amplitude to unity. The set of hopping phases encodes the magnetic flux $\Phi_j$ through each square lattice plaquette
\be \label{flux}
2\pi \frac{\Phi_j}{\Phi_0}=A_{j+\hat{x},j}+A_{j+\hat{x}+\hat{y},j+\hat{x}}-A_{j+\hat{x}+\hat{y},j+\hat{y}}-A_{j+\hat{y},j}
\ee
where $\Phi_0=h/e$ is the flux quantum.

In the presence of an external magnetic field, the non-interacting part of the Hamiltonian $\cH_0$ does not commute with translations by a single site. 
Therefore, the unit cell must be enlarged
into the so-called magnetic unit cell (MUC) that encloses an integer number of flux quanta \cite{Hofstadter1976}. The non-interacting part of the Hamiltonian $\cH_0$ commutes with translations defined by the operators
$T_{\vec{R}}=\sum_j \hat{c}^\dagger_{j+\vec{R}}\hat{c}_j$, where $\vec{R}$ is a translation vector corresponding to the MUC. It is possible to choose a vector potential such that $e^{iA}$ has spatial period $\vec{R}$ (see next paragraph), thus ensuring that
\be
[\cH_0,T_{\vec{R}}]=\sum_{j,\delta} \hat{c}^\dagger_{j+\vec{R}+\delta}\hat{c}_j \Big( e^{iA_{j+\delta,j}}-e^{A_{j+\vec{R}+\delta, j+\vec{R}}} \Big)
\ee
vanishes.
We choose the MUC lattice vectors $\vec{R}_1=M \vec{\hat{y}}$ and $\vec{R}_2=N \vec{\hat{x}}$,
with precisely $h/e$ flux through the MUC.

The phases $A_{j+\delta,j}$ are periodic only if the total magnetic flux through the MUC is zero. This can be ensured by performing a singular gauge transformation that inserts a point flux of $-h/e$
through a single square plaquette in the MUC so as to cancel the total flux. We emphasize that the insertion of the point flux, having magnitude $h/e$, has no effect on any physical observable, and is therefore a pure gauge transformation. To fix the gauge we impose the condition
$A_{j+\hat{x}+\hat{y},j+\unit{x}}=A_{j+\hat{x}}$ for all sites $j$, and the point flux is chosen to pass through the lower right plaquette of the MUC; with these conditions, the set of linear equations \eqref{flux} uniquely fix the gauge. See the supplementary information (SI) for an example of our gauge fixing scheme. ~\cite{SchirmerSM}.

To proceed with mean-field theory in real space, we consider an alternative Hamiltonian $\cH_{\rm MF}$ obtained by replacing the interaction term in $\cH$ by fermion bilinears coupled to fields.
The Hamiltonian, after the standard mean field decomposition in the pairing channel, is
\begin{equation} \label{eq7}
\begin{split}
\cH_{\rm MF}=&-\sum_{j,\delta}\Big(e^{iA_{j+\delta,j}}\hat{c}^\dagger_{j+\delta} \hat{c}_j+e^{iA_{j,j+\delta}}\hat{c}^{\dagger}_j \hat{c}_{j+\delta}\Big) -\mu \sum_j \hat{c}^\dagger_j \hat{c}_j\\ & - \sum_{j,\delta} \Big(\Delta_{j,\delta} \hat{c}^\dagger_{j+\delta}\hat{c}^\dagger_j+\Delta_{j,\delta}^*\hat{c}_{j}\hat{c}_{j+\delta} -\frac{|\Delta_{j,\delta}|^2}{U} \Big)
\end{split}
\end{equation}
where the vector field $\Delta_{j,\delta}$ is treated as a variational parameter whose optimal value is given by \cite{SchirmerSM}
\begin{equation} \label{selfcon}
\Delta_{j,\delta}=U\langle \hat{c}_j\hat{c}_{j+\delta}\rangle
\end{equation}
where the expectation value is with respect to the ground state of the mean-field Hamiltonian $\cH_{\rm MF}$.  
The Hamiltonian is invariant under the gauge transformation $\hat{c}_j\rightarrow e^{i\phi_j}\hat{c}_j$, $\Delta_{j,\delta}\rightarrow e^{i(\phi_{j+\delta}+\phi_j)}\Delta_{j,\delta}$, and $A_{j+\delta,j}\rightarrow A_{j+\delta,j}+\phi_{j+\delta}-\phi_{j}$.

We consider a lattice with $L\times L$ MUCs, with each MUC of size $M\times M$; we will take $L=10$ and $M=20$ for our numerical calculations.
We write the mean field Hamiltonian in momentum space by Fourier transforming
$\hat{c}_j=\frac{1}{L}\sum_{\vec{k}}e^{i\vec{k}\cdot\vec{R}}\hat{c}_\alpha(\vec{k})$
where $\vec{R}$ is the position vector for the origin of the MUC in which site $j$ lies and $\alpha$ is a site label within the MUC corresponding to site $j$. The number of allowed $\vec{k}$ magnetic momenta is $L^2$.
We have tested that our results do not change qualitatively when larger values of $L$ are used.

Assuming that the pairing field $\Delta_{j,\delta}$ has the same periodicity as the MUC,
we obtain the mean field Hamiltonian in the BdG form:
\begin{equation}
\cH_{\rm MF}=\frac{1}{2} \hat{C}^\dagger(\vec{k}) {\mathcal H}_{{\rm BdG}}(\vec{k}) \hat{C}(\vec{k})+\frac{1}{2}\sum_m\big(\epsilon_m(\vec{k})-\mu\big)+\sum_{j,\delta}\frac{|\Delta_{j,\delta}|^2}{U}
\end{equation}
where $\epsilon_m(\vec{k})$ are the eigen-energies of the non-interacting Hamiltonian $\cH_0$. The row vector $\hat{C}^\dagger(\vec{k}) =\big(\hat{c}^\dagger_1(\vec{k})~ \cdots~ \hat{c}^\dagger_{M^2}(\vec{k})~\hat{c}_1(-\vec{k})~\cdots~\hat{c}_{M^2}(-\vec{k}) \big)$  and
\begin{equation}
{\mathcal H}_{\rm BdG}(\vec{k})=
\begin{pmatrix}
h(\vec{k}) & \Delta_{\mathrm A}(\vec{k})\\
 \Delta_{\mathrm A}^\dagger(\vec{k})& -h^*(-\vec{k})
\end{pmatrix}
\end{equation}
where $\Delta_{\mathrm A}(\vec{k})=\Delta(\vec{k})-\Delta^T(-\vec{k})$ and $\Delta(\vec{k})$ is the Fourier transform of $\Delta_{j,\delta}$.
The BdG Hamiltonian ${\mathcal H}_{\rm BdG}(\vec{k})$ is $2M^2 \times 2M^2$, Hermitian, and so is diagonalizable by a unitary transformation. We write the quasiparticle operators, which are the eigenoperators of the Hamiltonian ${\mathcal H}$, as
\begin{equation} \label{qp ops}
\hat{\gamma}_m(\vec{k})=\sum_{\alpha} u_{\alpha,m}(\vec{k})\hat{c}_\alpha(\vec{k})+v_{\alpha,m}(\vec{k})\hat{c}^\dagger_\alpha(-\vec{k})
\end{equation}
where  $u_{\alpha,m}(\vec{k})$ and $v_{\alpha,m}(\vec{k})$ are determined from the eigenvectors of ${\mathcal H}_{\rm BdG}(\vec{k})$, and $m$ labels the BdG quasiparticle bands.
Note that the BdG Hamiltonian is particle-hole symmetric
\begin{equation}
{\mathcal P}{\mathcal H}_{\rm BdG}(\vec{k}){\mathcal P}^{-1}=-{\mathcal H}_{\rm BdG}(-\vec{k})
\end{equation}
where ${\mathcal P}=\tau_x {\mathcal K}$, $\mathcal K$ is the complex conjugation operator, and $\tau_x$ is a Pauli matrix acting on the particle and hole subspaces. This means that not all eigenvectors are independent -- if $\big(\vec{V}_1(\vec{k})~\vec{V}_2(\vec{k})\big)^T$ is an eigenvector of ${\mathcal H}_{\rm BdG}$ at positive $E(\vec{k})$, then $\big(\vec{V}_2(\vec{k})^*~\vec{V}_1(\vec{k})^*\big)^T$ is an eigenvector of ${\mathcal H}_{\rm BdG}$ at $-E(-\vec{k})$. This, in turn, implies that not all quasiparticle operators are independent
\begin{equation}
\hat{\gamma}_m(\vec{k})=\hat{\gamma}_{m'}^\dagger(-\vec{k})
\end{equation}
where band $m'$ is such that $E_m(\vec{k})=-E_{m'}(-\vec{k})$. Given all this, the full Hamiltonian $\hat{H}$ can be so expressed
\begin{equation}
\begin{split}
\cH_{\rm MF}=\sum_{\vec{k}}\sum_{m=1}^{M^2}\Bigg( & E_m(\vec{k})\hat{\gamma}^\dagger_{m}(\vec{k})\hat{\gamma}_{m}(\vec{k})+\frac{1}{2}\epsilon_m(\vec{k})-\frac{1}{2}E_m(\vec{k}) \Bigg) \\+ &\sum_{j,\delta}\frac{|\Delta_{j,\delta}|^2}{U},
\end{split}
\end{equation}
The ground state energy is given by
\begin{multline}\label{eq23}
E_g=\sum_{E_m(\vec{k})<0}E_m(\vec{k})+\frac{1}{2}\sum_{\vec{k}}\sum_{m=1}^{M^2}\Big(\epsilon_m(\vec{k})-\mu-E_m(\vec{k}) \Big)\\+ \sum_{j,\delta}\frac{|\Delta_{j,\delta}|^2}{U}.
\end{multline}
By inverting Eq.~\ref{qp ops}, we can express the expectation value $\langle \hat{c}_j\hat{c}_{j+\delta}\rangle$ in terms of  $u_{\alpha,m}(\vec{k})$ and $v_{\alpha,m}(\vec{k})$ (at zero temperature) to write the self-consistency equation \eqref{selfcon} as
\be
\Delta_{j,\delta}=\frac{U}{L^2}\sum_{\vec{k}}\sum_{m}e^{-i\vec{k}\cdot\Delta\vec{R}}u_{\alpha+\delta,m}(\vec{k})v^*_{\alpha,m}(\vec{k})
\ee
where $\alpha$ is the site within the MUC corresponding to site $j$ and $\Delta \vec{R}$ is difference in MUC location of site $j+\delta$ and site $j$. The subscript $\alpha+\delta$ refers to the nearest-neighbors of site $\alpha$, and is understood to be taken modulo the MUC. This equation is solved iteratively by starting with a random guess for the pairing potential $\Delta^{(0)}_{j,\delta}$, diagonalizing $\cH_{\rm BdG}$ to obtain $u_\alpha$'s and $v_{\alpha}$'s, and then obtaining a new pairing potential $\Delta^{(1)}_{j,\delta}$ using \eqref{selfcon}. The new pairing potential is treated as a new guess and the process is repeated until convergence is achieved to within a relative error of 10$^{-5}$. Many random initial guesses are tried, and the one which yields the lowest energy (expressed in \eqref{eq23}), is taken to be the ground state solution. Generically, a large majority of the initial guesses produce the ground state. Solutions with higher energies than the ground state are taken to be excited states. We emphasize that the solutions to the self-consistency equation are gauge invariant in the sense that if $\Delta_{j,\delta}$ is a solution to the self-consistency equations with energy $E_g$ in a certain gauge, 
then $e^{i(\phi_{j+\delta}+\phi_j)}\Delta_{j,\delta}$ is a self-consistent solution in another gauge, related to the first by the gauge transformation $\phi_j$, with the same energy $E_g$.

The BdG Hamiltonian $\cH_{\rm BdG}(\vec{k})$, describing a system in two spatial dimensions and being particle-hole symmetric only, resides in class D of the Altland-Zirnbauer classification \cite{altland1997}, and is therefore characterized by a $\mathbb{Z}$ bulk topological invariant, called the Chern number $\mathcal{C}$ \cite{Thouless1982}. For superconducting systems, the Chern number $\mathcal{C}$ counts the number of chiral Majorana edge modes \cite{majorana1937,Wilczek2009,qi2010chiral,leijnse2012} and, in particular when $\mathcal{C}$ is odd, the system is topologically
distinct from an ordinary quantum Hall state which, in the BdG framework, possesses an even Chern number. The Chern number of the BdG Hamiltonian can be computed in the same fashion as for non-interacting Hamiltonians: one defines the Berry connection, determined by the eigenvectors $|u^m(\vec{k})\rangle$ of the BdG Hamiltonian $\cH_{\rm BdG}(\vec{k})$, where $m$ labels the band, as
\begin{equation}
A^{mn}_{\mu}(\vec{k})=i\langle u^m(\vec{k})| \partial_{\mu} |u^n(\vec{k})\rangle
\end{equation}
where $\partial_{\mu}$ is the shorthand for $\partial/\partial{k_{\mu}}$. The Berry curvature is then defined as
\begin{equation}
 F^{mn}_{\mu\nu}(\vec{k})=\partial_{\mu} A^{mn}_{\nu} - \partial_{\nu} A^{mn}_{\mu} + i \left[{A}_{\mu}, {A}_{\nu} \right]^{mn}
\end{equation}
The only non-zero components of the Berry curvature are $F_{\rm xy}=-F_{\rm yx}\equiv F$. The Chern number is then given by the Berry curvature integrated over the magnetic Brillouin zone (MBZ):
\begin{equation} \label{hall_formula}
\mathcal{C} = \frac{1}{2\pi} \int_{\rm MBZ} d^2k  {\rm Tr} \left[ { F}(\vec{k}) \right]_{E(\vec{k})<0}
\end{equation}
We compute Eq. \ref{hall_formula} numerically using the method by Fukui, et. al \cite{fukui2005} -- the Berry curvature is determined on a grid in a discretized MBZ by defining

\begin{equation}
M^{mn}_\lambda(\vec{k_{\alpha}})=\langle u^m (\vec{k_{\alpha}})| u^n (\vec{k_{\alpha}}+\vec{e}_{\lambda}) \rangle \label{eq19}
\end{equation}
 The points on the grid are labeled by $\vec{k_{\alpha}}$ and the spacing vectors are $\vec{e}_{\lambda}$ where $\lambda=1,2$. In terms of the link variables defined as
\begin{equation}
U_\lambda(\vec{k_\alpha})=\frac{\det M_\lambda(\vec{k_\alpha})}{|\det M_\lambda(\vec{k_\alpha})|}
\end{equation}
 the discrete Berry curvature at each point on the grid is given by
\begin{multline}
\tilde{F}(\vec{k_{\alpha}})= \ln\Big( U_1(\vec{k_{\alpha}})  U_2(\vec{k_{\alpha}}+\vec{e_1})  \\ U_1^{-1}(\vec{k_{\alpha}}+\vec{e_2}) U_2^{-1}(\vec{k_{\alpha}}) \Big)
\end{multline}
The Chern number is then given by
\begin{equation} \label{eq22}
\mathcal{C}=\frac{1}{2\pi i}\sum_{\vec{\alpha}}\tilde{F}(\vec{k_{\alpha}})
\end{equation}

\section*{Results}

\begin{figure*}
\includegraphics[width=\textwidth]{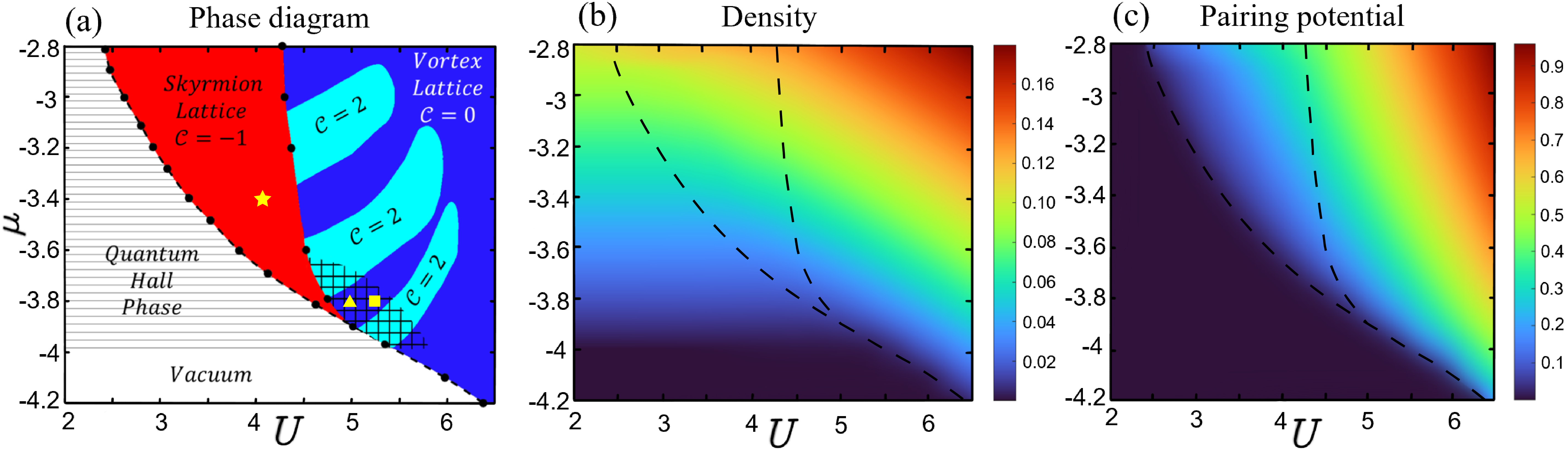}
\caption{(a) The mean-field phase diagram as a function of the magnitude of the nearest neighbor attractive interaction $U$ and the chemical potential $\mu$, in units where the hopping amplitude is set to unity. Horizontal black lines in the Quantum Hall Phase indicate the energies of the Landau levels. The phase with skyrmion lattice region is marked in red. The regions with vortex lattice are indicated in blue and cyan, and the gridded region has competing vortex lattice/dimer vortex lattice phases. The yellow star, square, and triangle corresond to the skyrmion, vortex lattice, and dimer vortex lattice phases shown in Fig. 3, respectively. (b) The average particle number per site (density in units of $a^{-2}$ where $a$ is the lattice constant) as a function of $\mu$ and $U$. (c) Spatially averaged magnitude of the pairing potential as a function of $\mu$ and $U$. Note that the discontinuous changes in the spatially averaged pairing potential and the average density at the (first order) phase transition are too small to be visible in the plots; for example, for $\mu=-3.6$, at the vortex lattice - skyrmion lattice phase boundary, the former is 0.006 (reflecting a 4\% change in the magnitude) and the latter is 0.0002 (0.5\%).} \label{diagram}
\end{figure*}

We are primarily interested in the ground state and the low-energy fermionic spectrum of the model. Indeed we find a rich phase diagram with unexpected ground states, as shown in Fig. \ref{diagram}, which we discuss in detail below.

\subsection{Quantum Hall phase}

Superconducting regions in the phase diagram are denoted with color, and the non-superconducting region is indicated in white, where the single particle spectrum forms very flat and nearly equally spaced Landau level-like Chern bands (we will henceforth refer to them as LLs), with superconducting Chern number $\mathcal{C}=-2\nu$ where $\nu$ is the filling factor \cite{aidelsburger2015artificial}. The energy of these bands is indicated by the horizontal black lines in the figure, and this phase is deemed a QH phase. 

It should be noted that if the chemical potential is tuned so as to lie precisely within a LL, superconductivity extends to arbitrarily low interaction strength. The indicated boundary between the QH and the superconducting regions (dashed line) is determined for chemical potentials that lie between LLs.

\subsection{Skyrmion phase}

Starting from the non-superconducting region and increasing the attractive interaction strength, the QH system transitions into a superconducting ground state which forms a skyrmion lattice \cite{takigawa2001,li2009skyrmion,garaud2012,garaud2015,Becerra2015,zhang2016} (discussed below at more length) for values of the chemical potential corresponding to filling factor $\nu \gtrsim 3$. This region, where the SC Chern number $\mathcal{C}=-1$, is indicated in red in Fig. \ref{diagram}.

\begin{figure*}[h]
\includegraphics[width=\textwidth]{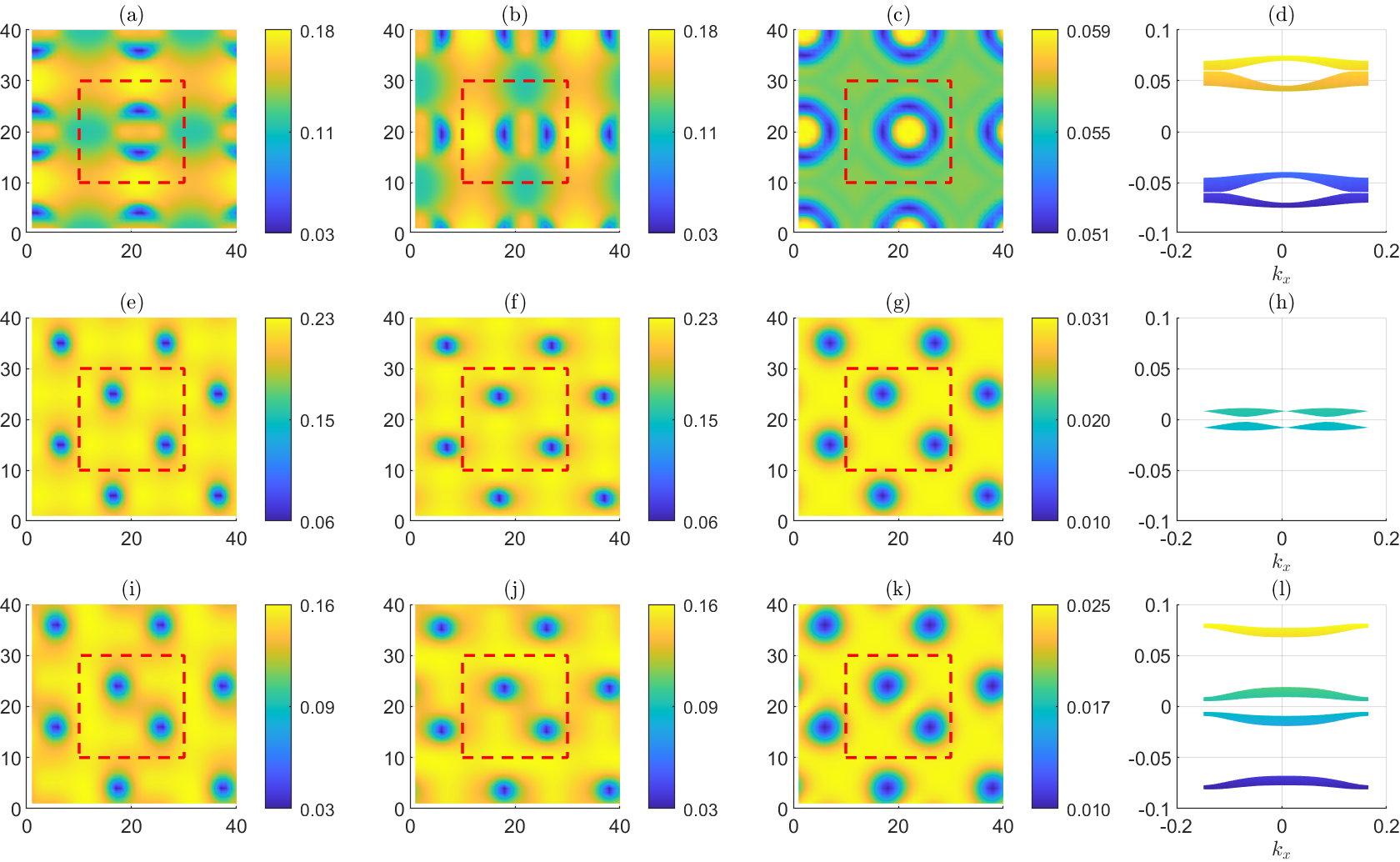}
\caption{ (a-b)  The magnitude of the components of the pairing potential $|\Delta_{j,x}|$ and $|\Delta_{j,y}|$, (c) the particle density, (d) the low energy quasiparticle bands for the skyrmion lattice solution with $U=4.05$ and $\mu=-3.4$. Panels (e)-(h) show the same but for the vortex lattice solution with  $U=5.25$ and $\mu=-3.8$. Panels (i)-(l) show the same but for the dimer lattice solution with $U=5$ and $\mu=-3.8$. The red dashed lines denote the boundary of the MUC.} \label{fig1}
\end{figure*}

In a chiral p-wave superconductor, a skyrmion is a closed domain of inverted chirality with half-quantum vortices (HQVs) located on the domain wall \cite{takigawa2001,li2009skyrmion,garaud2012,garaud2015,Becerra2015,zhang2016}. As shown below, these configurations have a skyrmionic texture in a suitable pseudospin representation and carry a non-zero topological charge. The magnitudes of the two components of the order parameter, $|\Delta_{j,x}|$ and $|\Delta_{j,y}|$, the particle density, and the low energy BdG quasiparticle band structure are shown in Fig. 3 (a)-(d). It can be seen that there are two vortices per component of the order parameter which each have different locations (the vortices are {\it coreless}). The phase of only one component of the order parameter winds by $2\pi$ around each vortex while the other component remains constant. The vortices are HQVs since each is associated with a $h/4e$ magnetic flux (the entire skymion in each MUC is associated with $h/e$ flux). The skyrmion forms a ring structure in the density, as seen in 2 (c), which may provide a signature of skyrmions in scanning tunneling microscopy (STM) measurements. The fermionic spectrum in the skyrmion phase, shown in Fig. 2 (d), is gapped, indicating the absence of zero-energy Majorana bound states in the bulk. The SC Chern number, computed using \eqref{hall_formula} is $\mathcal{C}=-1$, signifying that it is in the phase of p-wave superconductors with non-Abelian excitations and has exactly one chiral Majorana edge mode.

It is useful to contrast the skyrmion phase from the more familiar vortex phase, which is discussed below. Unlike in the skyrmion phase, the vortices in each component occur at the same location in the vortex phase. Also, in the vortex phase the 
``total gap'' $\sqrt{|\Delta_{j,x}|^2 + |\Delta_{j,y}|^2}$ almost vanishes at the locations of the vortices; for the skyrmion lattice, in contrast, the total gap is everywhere finite. Finally, the vortex phase has a low energy Majorana band, which is absent in the skyrmion phase.

In order to classify the self-consistent solutions of the order parameter $\Delta_{j,\delta}$, it is convenient to define a topological invariant called the (real space) skyrmion number
\be \label{skynum}
\mathcal{Q}=\frac{1}{4\pi}\int \dd^2 \vec{r}~ \unit{n}\cdot \big( \partial_x \unit{n} \times \partial_y \unit{n} \big)
\ee
where the pseudospin vector $\unit{n}$ is defined in terms of the superconducting order parameter in the continuum
$\hat{n}_\alpha=\frac{\vec{\Delta}^\dagger \sigma_\alpha \vec{\Delta}}{\vec{\Delta}^\dagger \vec{\Delta}}$, where
$\vec{\Delta}= (\Delta_x(\vec{r}),  \Delta_y(\vec{r}))^T$
and $\sigma$ is a Pauli matrix ($\alpha=x,y,z$). We expect $\mathcal{Q}=0$  for vortex solutions and $\mathcal{Q}=2$ for skyrmions \cite{garaud2012}. On the lattice, a problem arises with this definition of $\mathcal{Q}$. The order parameter transforms non-locally under a gauge transformation $\hat{c}_j\rightarrow e^{i\phi_j}\hat{c}_j$, i.e. $\Delta_{j,\delta}\rightarrow e^{i(\phi_{j+\delta}+\phi_j)}\Delta_{j,\delta}$, and so the vector $\unit{n}(\vec{r}_j)$ transforms as
$\unit{n}(\vec{r}_j)\rightarrow R_{\mathrm{xy}}(\delta \theta)\unit{n}(\vec{r}_j)$,
where $R_{\mathrm{xy}}(\delta \theta)$ is a rotation matrix in the xy-plane by an angle $\delta \theta =\phi_{j+\hat{y}}-\phi_{j+\hat{x}}$. Consequently, the skyrmion number $\mathcal{Q}$ defined in this way is not invariant under gauge transformations on the lattice (the skyrmion number in the continuum, in contrast, is gauge invariant since there the continuum order parameter transforms locally under gauge transformations).

To construct a gauge invariant skyrmion number on the lattice, we
define the {\it gauge invariant} phase difference between the components of the order parameter in the following way. We first define the phase of the two order parameters $\theta_{j,\delta}$
\be
\Delta_{j,\delta}=|\Delta_{j,\delta}|e^{i\theta_{j,\delta}}
\ee
where $\delta=x,y$. Instead of the bare phase, we consider the phase $\tilde{\theta}_{j,\delta}=\theta_{j,\delta}-A_{j+\delta,j}$ which by itself is not gauge invariant, but the combination $\tilde{\theta}_{j,y}-\tilde{\theta}_{j,x}$ $\it{is}$ gauge invariant:
\be
\begin{split}
\tilde{\theta}_{j,y}-\tilde{\theta}_{j,x} \rightarrow &\theta_{j,y}+\phi_j+\phi_{j+\hat{y}}-\big( A_{j+\hat{y},j}+\phi_{j+\hat{y}}-\phi_{j} \big) \\ - &\theta_{j,x}-\phi_j-\phi_{j+\hat{x}}+\big( A_{j+\hat{x},j}+\phi_{j+\hat{x}}-\phi_{j} \big) \\
=&\big( \theta_{j,y}-A_{j+\hat{y},j} \big) - \big(\theta_{j,x} - A_{j+\hat{x},j} \big) \\
= &\tilde{\theta}_{j,y}-\tilde{\theta}_{j,x}
\end{split}
\ee
 This motivates us to define a pseudospin vector on the lattice using $\tilde{\theta}_{j,\delta}$
 \be
 \begin{split}
 \hat{n}_{j,\alpha}=\frac{\vec{\Delta}_j^\dagger \sigma_\alpha \vec{\Delta}_j}{\vec{\Delta}_j^\dagger \vec{\Delta}_j}~~~ \mathrm{where}~~~ \vec{\Delta}_j=\begin{pmatrix}
|\Delta_{j,x}|e^{i\tilde{\theta}_{j,x}} \\
|\Delta_{j,y}|e^{i\tilde{\theta}_{j,y}}
\end{pmatrix}
 \end{split}
 \ee
 The pseudospin $\hat{n}_{j,\alpha}$ is gauge invariant on the lattice and the skyrmion number, computed using a lattice-discretized version of \eqref{skynum}, is $\mathcal{Q}=2$ for a skyrmion and $\mathcal{Q}=0$ for two vortices (both evenly spaced and dimer vortices discussed below) when rounded to the nearest integer. This is consistent with the values found in continuum systems \cite{garaud2012,garaud2015,Becerra2015}. Furthermore, it can be shown that the lattice pseudospin vector is identical to the continuum pseudospin vector, defined above, in the continuum limit. Therefore we conclude that the 
deviation of the lattice skyrmion number
 from an integer is a result of lattice discretization; using a finer lattice (i.e. more sites per MUC) will yield skyrmion numbers closer to their continuum limit values.

\begin{figure} 
\includegraphics[width=.5\textwidth]{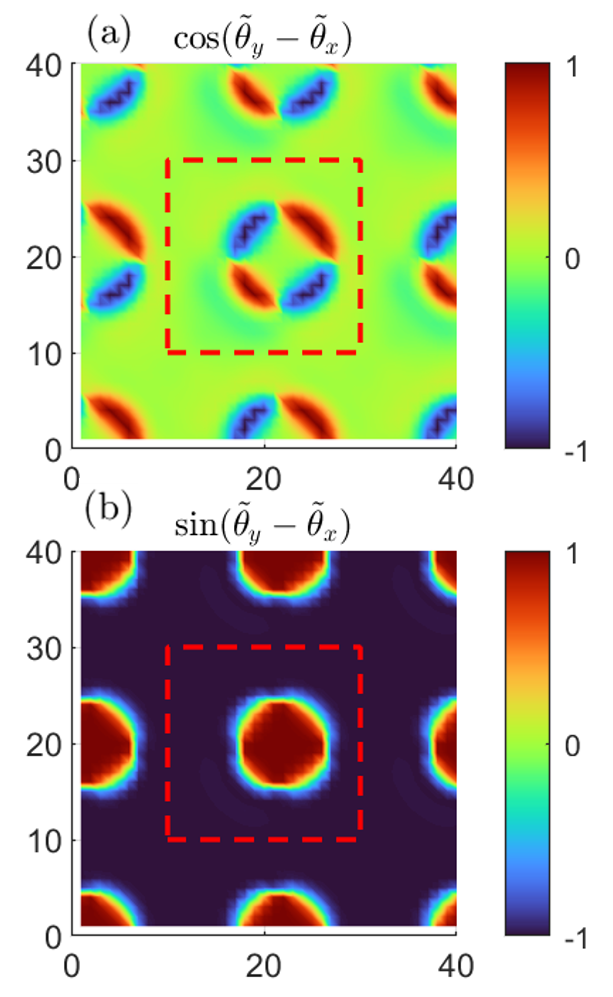}
\caption{(a) The cosine of the gauge invariant phase difference between the $x$ and $y$ components of the order parameter for the skyrmion phase. Along the boundary of the skyrmion, the pairing symmetry oscillates between $p_x+p_y$ and $p_x-p_y$. (b) The sine of the gauge invariant phase difference between the $x$ and $y$ components of the order parameter for the skyrmion phase. Inside the skyrmion, the pairing symmetry is $p_x+ip_y$, and outside, the pairing symmetry is $p_x-ip_y$.} \label{fig2}
\end{figure}

It is also instructive to look at the spatial variation of the (gauge invariant) phase difference $\tilde{\theta}_{j,y}-\tilde{\theta}_{j,x}$ in the skyrmion phase. Figure \ref{fig2} shows the cosine and sine of the angle difference for the skyrmion phase. The cosine shows alternating sign between the half-quantum vortices, a signature of skyrmions in $p$-wave superconductors \cite{Becerra2015}. The sine plot displays another defining feature of skyrmions in this context -- closed domains of inverted chirality. In the present case, the pairing symmetry is $p_x+ip_y$ (the phase difference is $\pi/2$) inside the skyrmion, and outside it is $p_x-ip_y$ (the phase difference is $-\pi/2$). The half-quantum vortices reside on this domain wall. We note that the phase difference plots for the vortex and dimer vortex phase are comparatively featureless --  the phase difference is pinned to $\pi/2$, implying that the entire system has $p_x+ip_y$ symmetry. 

\subsection{Vortex lattice and dimerized-vortex lattice}

The region where the ground state forms a vortex lattice is indicated by blue and cyan in Fig. \ref{diagram}, with SC Chern numbers equal to $\mathcal{C}=0$ and $\mathcal{C}=2$, respectively. The vortex lattice solution is energetically favored in the superconducting region when the chemical potential is low or the interaction strength is high. Note that for low chemical potentials the transition is directly from the QH phase is into the vortex lattice phase.

The magnitudes of the two components of the order parameter, $|\Delta_{j,x}|$ and $|\Delta_{j,y}|$, the particle density, and the low energy BdG quasiparticle band structure for the vortex lattice phase are shown in Fig. 3 (e)-(h). The vortices in each component occur at the same location and form a square lattice. Note that the ``total gap'' $\sqrt{|\Delta_{j,x}|^2 + |\Delta_{j,y}|^2}$ almost vanishes at the locations of the vortices. The vortex lattice phase has a low-energy band structure (shown in \label{fig1} (h)) which is consistent with the findings of previous studies of Majorana bands \cite{grosfeld2006electronic,cheng2009,PhysRevB.82.094504,Ludwig2011,kraus2011majorana,Lahtinen2012,laumann2012disorder,zhou2013hierarchical,biswas2013majorana,silaev2013majorana,murray2015,PhysRevB.92.134519,ariad2015effective}, where the dynamics was described by Majoranas hopping between vortices with nearest neighbor and next-nearest neighbor hopping. The spectrum has a small gap and we find that the Chern number is $\mathcal{C}=0,2$ in the vortex phase, which is also consistent with previous studies \cite{murray2015,chaudhary2020}. This supports the claim that the non-Abelian phase is not possible for a lattice of vortices with one vortex per vortex unit cell \cite{murray2015,Mishmash2019,chaudhary2020}.

Additionally, we find that vortices show a tendency to dimerize (i.e. form a lattice with two vortices per vortex unit cell). This phase is energetically competitive  in the black gridded region of the phase diagram; for some parameters it is the clear ground state whereas in other cases it is only slightly higher in energy to the vortex lattice solution. It is also found as a higher energy self-consistent solution in much of the phase diagram. The magnitudes of the two components of the order parameter, $|\Delta_{j,x}|$ and $|\Delta_{j,y}|$, the particle density, and the low energy BdG quasiparticle band structure for the dimer vortex lattice phase are shown in 2 (i)-(l). As can be seen, the vortices are not equally spaced. This breaks the vortex magnetic translation symmetry \cite{jeon2019,Mishmash2019} and allows for odd SC Chern number and hence the non-Abelian phase. We find that the SC Chern number $\mathcal{C}=1$ is the same as that of the background $p_x+ip_y$ superconductor. It is encouraging to find that the system breaks the vortex translation symmetry all on its own; translation symmetry-breaking perturbations may thus not be required to realize the non-Abelian phase of vortex lattices in $p$-wave superconductors. \cite{kraus2011majorana,laumann2012disorder,Mishmash2019}.

\vspace{4 mm}

\begin{figure*} 
\includegraphics[width=\textwidth]{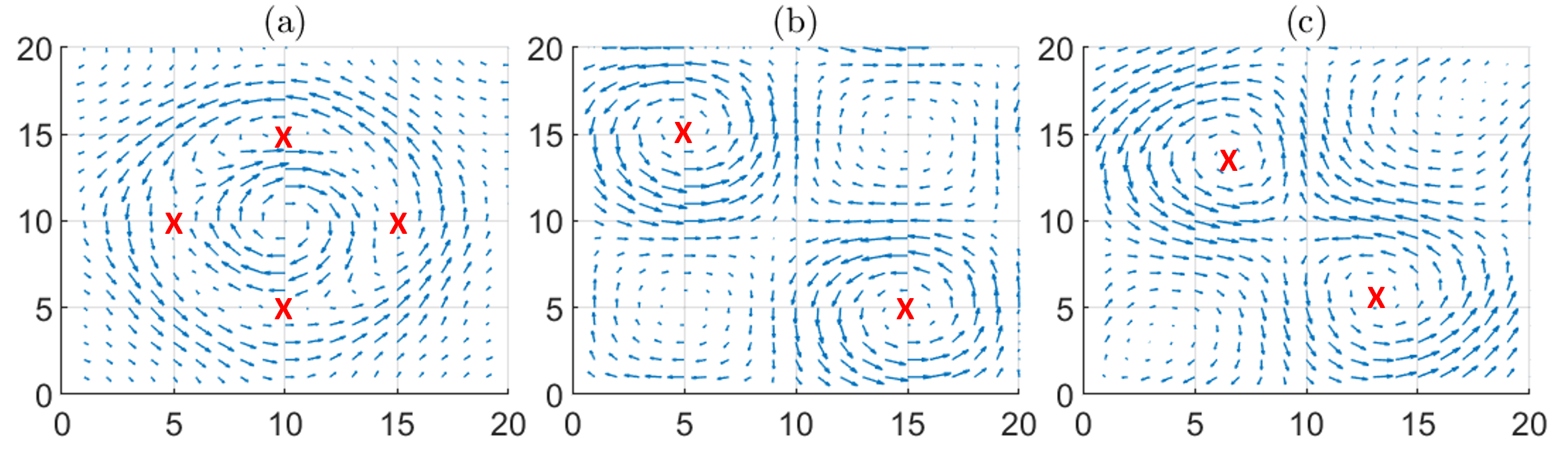}
\caption{The current texture for a skyrmion (a), a pair of vortices in a vortex lattice (b), and pair of vortex dimers (c). Half-vortices in (a) and vortices in (b), (c) are marked by {\bf {\color{red} X}}.} \label{fig3}
\end{figure*}

The current texture for the skyrmion, vortex, and dimer vortex lattice phases are shown in Fig. \ref{fig3}. The current is determined by evaluating the expectation value of the current operator
\be
\hat{J}_{j,\delta}=i \Big(e^{iA_{j+\delta,j}}\hat{c}^\dagger_{j+\delta}\hat{c}_j - e^{iA_{j,j+\delta}}\hat{c}^\dagger_j \hat{c}_{j+\delta} \Big)
\ee
in the ground state. This expression can be derived by considering variation of the Hamiltonian with respect to the hopping phases, or by using the continuity equation \cite{watanabe2019}. The current texture for a skyrmion (shown in Fig. \ref{fig3}a) shows inner and outer currents flowing in opposite directions. This structure can be understood as coming from the combined supercurrent of the bound half-quantum vortices. The vortex lattice phase (\ref{fig3}b) shows relatively separated current circulation for each vortex whereas for the vortex dimers  (\ref{fig3}c), the current percolates between the vortices in the dimer.

The spatially averaged pairing potential and the average density are also shown in Fig.~\ref{diagram}. The 
discontinuous jumps in these quantities at the (first order) phase boundaries are small because they arise primarily from the regions in the immediate vicinity of the vorcies and the skyrmions. For example, for $\mu=-3.6$, at the vortex lattice - skyrmion lattice phase boundary, the change in the spatially averaged pairing potential is is 0.006 (reflecting a 4\% change in the magnitude) and that in the average density is 0.0002 (0.5\% change in the magnitude). These jumps are not visible in Fig.~\ref{diagram} except for the change in the pairing potential from the quantum Hall to vortex / skyrmion lattice phase.

\section*{Discussion}

In spite of the attention that vortices have received in the context of topological $p$-wave superconductivity in QH systems, the question of whether or not the vortex lattices are the stable self-consistent ground state solutions has been relatively unexplored. Nevertheless, recent studies have pointed out that systems with a regular lattice of Abrikosov vortices carrying flux $h/2e$ have an even SC Chern number $\mathcal{C}$, due to the vortex magnetic translation symmetry which enforces an even number of Dirac crossings, and hence these systems have the same classification as the QH phase with fermionic (as opposed to Majorana) chiral edge modes \cite{murray2015,Mishmash2019,chaudhary2020}. These results are indeed consistent with ours, for the portion of the phase diagram at relatively strong pairing strength, but we find cases where the system seems to circumvent this problem by forming skyrmions or vortex dimers, and thereby allowing for odd SC Chern number and hence chiral Majorana edge modes. 

Some insight into the origin of the vortex dimer phase can be gained by analogy to Peierls-type physics: a Peierls distortion opens a larger gap, pushing the states near zero energy to lower energies, as shown in Figs. 2 (d), (h), (l), thereby reducing the energy of the Majorana band. However, whether such distortion actually occurs depends on other bands as well as the last term in Eq.~\ref{eq23}. Similarly, the formation of skyrmions also opens a large gap at the Fermi energy which can stabilize it in some parameter regimes.

It should be noted that our choice of a square MUC containing one quantum of flux $h/e$, which is limited by the need of computational resources, 
forces the skyrmions, Abrikosov vortices, and vortex dimers to all form a square lattice. Choosing a larger number of magnetic flux quanta per MUC would constrain the lattice to a lesser degree and could allow for a triangular lattice of vortices. Although the lattice type may change, we expect the phase diagram to qualitatively remain the same, especially when the skyrmions, vortices, and vortex dimers are far separated.

A comment on the center of mass (COM) magnetic momentum of the Cooper pairs is also in order. Our assumption that the mean field pairing potential $\Delta_{j,\delta}$ has the same periodicity as the MUC of the non-interacting system (described by the Hamiltonian $\cH_0$) only allows for pairing between fermions of opposite magnetic momenta $\vec{k}$ and $-\vec{k}$, resulting in Cooper pairs with zero COM magnetic momentum. (A non-zero COM momentum would manifest through a phase twist across an MUC.) 
In studies of the SC Hofstadter model in large magnetic fields, pairing instabilities can also occur at finite COM momentum, due to degeneracies arising from the magnetic translation symmetry, leading to Fulde-Ferrell-Larkin-Ovchinnikov (FFLO) superconductivity \cite{zhai2010,PhysRevLett.104.255303,Sohal2020,shaffer2021}. We neglect in our work FFLO pairing because our interest has primarily to do with skyrmion / Majorana physics in the continuum limit, and also because FFLO pairing has not yet been decisively observed in condensed matter systems.

We end with a remark on the experimental relevance of our results. We expect our model with nearest neighbor attractive interaction to be relevant to $p$-wave superconductors under external magnetic fields. In addition, our study will provide a guidance to understand physics in  $s$-wave superconductors with strong spin orbit coupling under external magnetic field, or in heterostructures consisting of a 2D SC coupled to a 2D QH system. 

\section*{Acknowledgements}We are grateful to Egor Babaev and Mohit Randeria for enlightening discussions. JS thanks Prachi Singh for help with figures. JS and JKJ were supported in part by the U. S. Department of Energy, Office of Basic Energy Sciences, under Grant no. DE-SC-0005042. We acknowledge Advanced CyberInfrastructure computational resources provided by The Institute for CyberScience at The Pennsylvania State University. JS and CXL acknowledge the partial support from New Initiative research grant (grant No. KA2018-98553) of the Pittsburgh Foundation. CXL also acknowledges the partial support from the office of Naval Research (Grant No. N00014-18-1-2793) and NSF through the Princeton University’s Materials Research Science and Engineering Center DMR-2011750. 
\newpage
\bibliography{../../../References/my}
\newpage

\onecolumngrid
\section{Supplementary Materials}
\subsection{An example of periodic gauge choice}
\begin{figure}[H] 
\begin{center}
\includegraphics[width=0.5\textwidth]{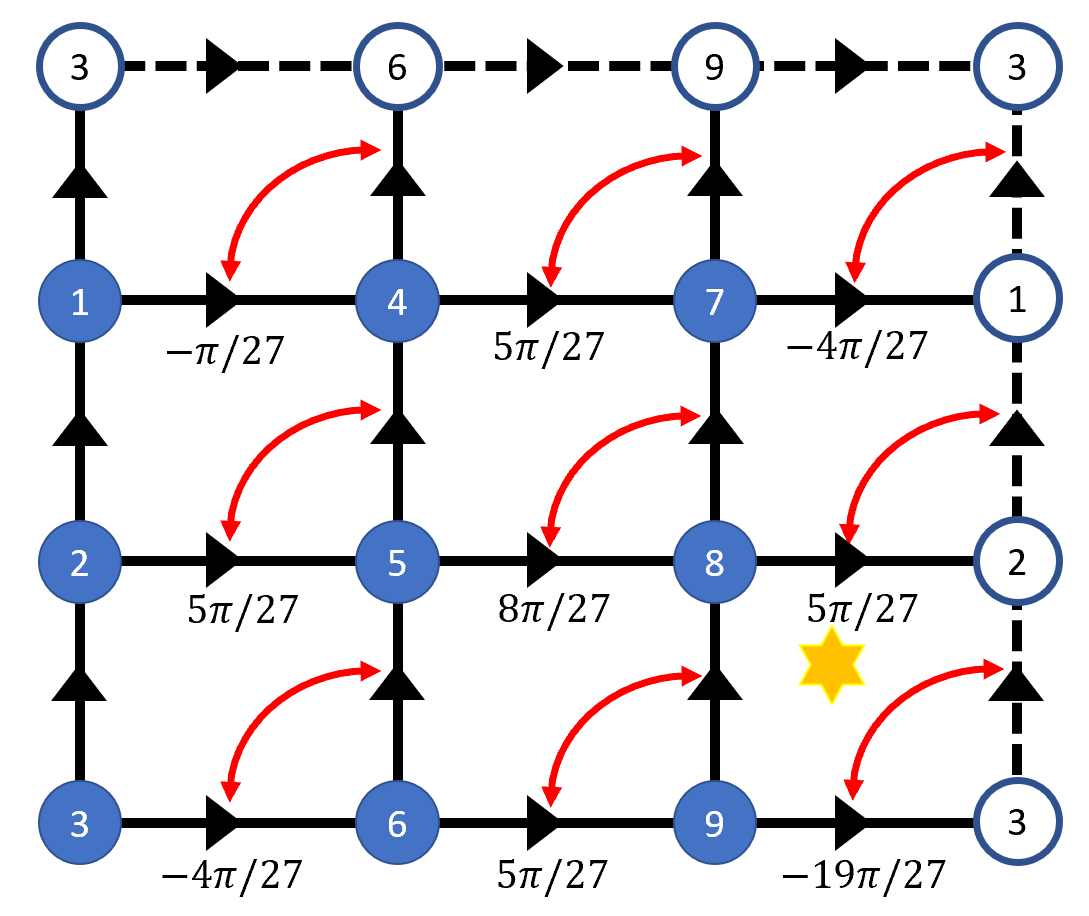}
\end{center}
\caption{An example of the gauge choice for a $3 \times 3$ MUC, with flux $\Phi_0/9$ passing through each plaquette. The MUC labels are given on each site. The hopping phases $A_{j+\delta,j}$ are given for the horizontal edges within the MUC; edges connected by red arrows have the same hopping phases (note the directionality of the edges). The dashed edges on the right are the periodic images of the edges on the far left. Similarly, the dashed edges on the top are periodic images of the edges on the very bottom. The opposing $\Phi_0$ flux is inserted into the lower rightmost square plaquette (the one marked by the star), which has a flux of $-8\Phi_0/9$ passing through it. The sum of the hopping phases around any closed path enclosing flux $\Phi$ is equal to $2\pi \Phi/\Phi_0$.} \label{S1}
\end{figure}

In Fig. \ref{S1}, we illustrate the gauge choice discussed in the main text by giving an example for a small magnetic unit cell (MUC). Here the MUC is of size $3\times3$ and the flux through each plaquette is $\Phi=\Phi_0/9$ ($\Phi_0=h/e$ is the flux quantum), except for the lower rightmost plaquette (marked by a star) where an additional point flux has been inserted resulting in a net flux of $\Phi_0/9-\Phi_0=-8\Phi_0/9$. The values of the hopping phases $A_{j+\delta,j}$ are given for the horizontal edges of the lattice within the MUC. Hopping phases along edges connected by red arrows are set equal -- this is the condition $A_{j+\unit{x}+\unit{y},j+\unit{x}}=A_{j+\unit{x},j}$ (see main text). The phases repeat periodically throughout the system.

\subsection{Excited states}
In addition to the ground state solutions discussed in the main text, we find higher energy self-consistent vortex lattice and skyrmion lattice solutions. 

\begin{figure}[H]
\includegraphics[width=\textwidth]{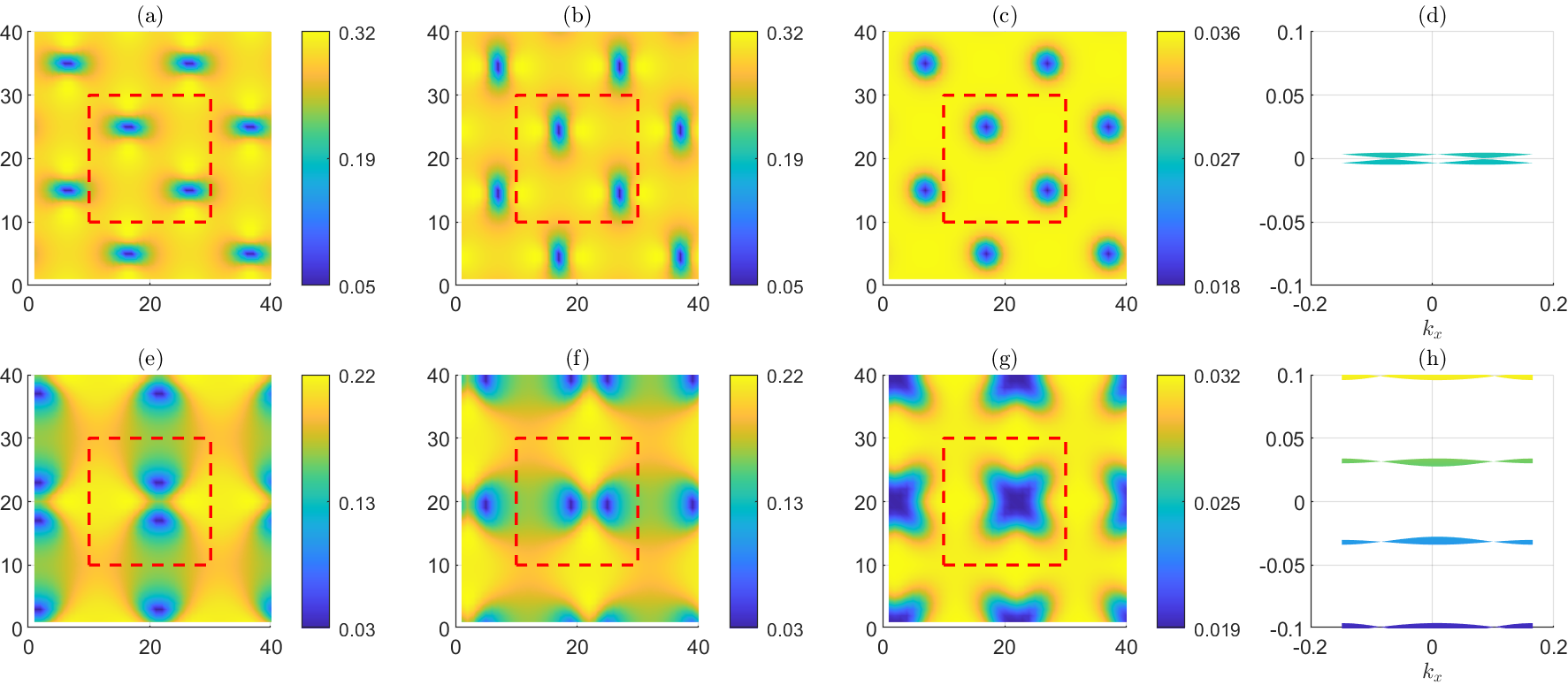}
\caption{(a-b)  The magnitude of the components of the pairing potential $|\Delta_{j,x}|$ and $|\Delta_{j,y}|$, (c) the particle density, (d) the low energy quasiparticle bands for the excited state vortex lattice solution with $U=5.5$ and $\mu=-3.8$. Panels (e)-(h) show the same but for the excited state skyrmion lattice solution with $U=5$ and $\mu=-3.75$. The red dashed lines denote the boundary of the MUC.} \label{S2}
\end{figure}

The excited vortex lattice solution is found to have $p_x-ip_y$ symmetry, instead of $p_x+ip_y$ which is found for the ground state vortex and dimer vortex lattices. The magnitudes of the two components of the order parameter, $|\Delta_{j,x}|$ and $|\Delta_{j,y}|$, the particle density, and the low energy BdG quasiparticle band structure are shown in  Fig. \ref{S2} (a)-(d). The profile of each vortex is less isotropic than that of $p_x+ip_y$ vortices as can be seen in Fig. \ref{S2} (a) and (b) (also cf. Fig. 2 (e) and (f) from the main text). The spectrum displays low energy hybridized Majorana bands, and the SC Chern number is $\mathcal{C}=0,-2$.

\begin{figure}[H] 
\begin{center}
\includegraphics{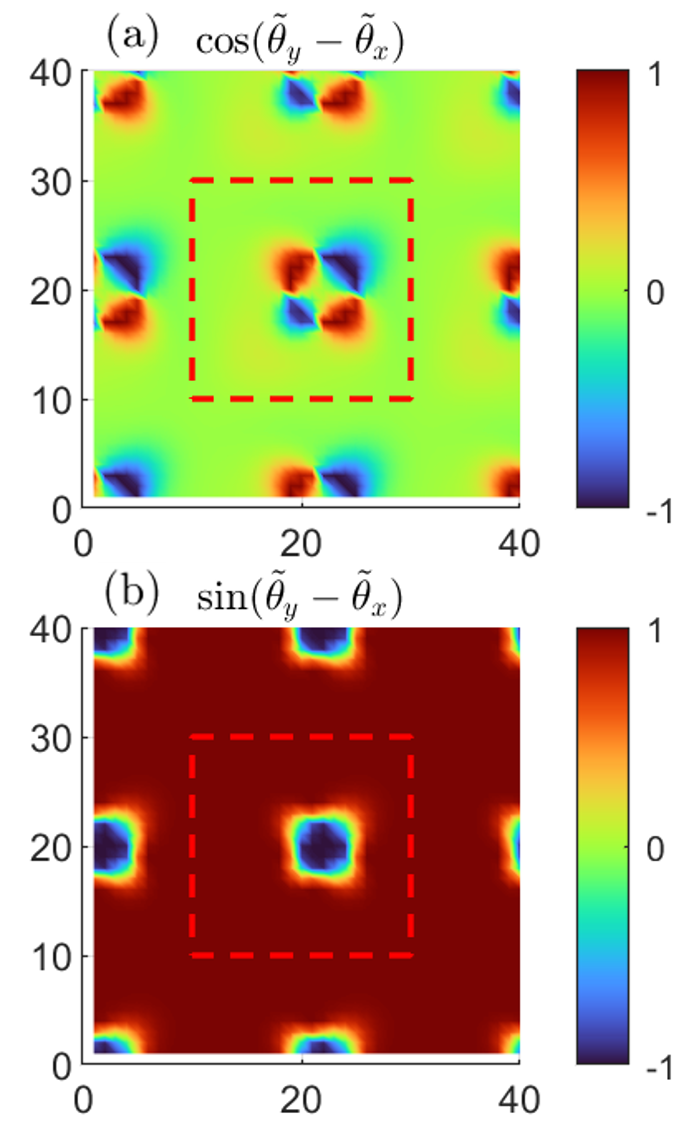}
\end{center}
\caption{(a) The cosine of the gauge invariant phase difference between the $x$ and $y$ components of the order parameter for the excited state skyrmion phase. Along the boundary of the skyrmion, the pairing symmetry oscillates between $p_x+p_y$ and $p_x-p_y$. (b) The sine of the gauge invariant phase difference between the $x$ and $y$ components of the order parameter for the excited state skyrmion phase. Inside the skyrmion, the pairing symmetry is $p_x-ip_y$, and outside, the pairing symmetry is $p_x+ip_y$.}\label{S3}
\end{figure}

Furthermore, we find skyrmions, which have their chirality inverted relative to the ground state skyrmions, among the excited states.  The magnitudes of the two components of the order parameter, $|\Delta_{j,x}|$ and $|\Delta_{j,y}|$, the particle density, and the low energy BdG quasiparticle band structure are shown in  \label{fig4} (e)-(h). The order parameters and the particle density look qualitatively different from the ground state skyrmions but the quasiparticle spectrum is still gapped. The SC Chern number in this phase is $\mathcal{C}=-1$. Fig. \ref{S3} shows the cosine and sine of the gauge invariant relative phase. The cosine (Fig. \ref{S3}a) shows the characteristic alternating sign along the edge of the skyrmion. Inside the skyrmion, the pairing symmetry is $p_x-ip_y$ and outside the skyrmion the pairing symmetry is $p_x+ip_y$, as seen in Fig. \ref{S3}b, which is the opposite of that of the ground state skyrmions discussed in the main text.

\newpage

\newpage

\newpage

\newpage

\end{document}